\documentclass[aps,prd]{revtex4}

\begin{document}
% You should use BibTeX and revtex.bst for references
\bibliographystyle{revtex}

\preprint{MADPH-01-1245, P3-17, hep-ph/0110242}
\begin{flushright}
MADPH-01-1245 \\
P3-17 \\
hep-ph/0110242 \\
\end{flushright}

\title{Physics of the radion in the Randall-Sundrum scenario}

\author{Graham D. Kribs}
\email[]{kribs@pheno.physics.wisc.edu}
%\thanks{}
\affiliation{Department of Physics, University of Wisconsin, 
             Madison, WI 53706}


\date{\today}

\begin{abstract}
In the Randall-Sundrum solution to the hierarchy problem, 
the fluctuations of the size of the extra dimension are
characterized by a single scalar field, called the radion. 
The radion is expected to have a mass somewhat lower than the TeV scale 
with couplings of order 1/TeV to the trace of the energy momentum tensor.  
In addition, the radion can mix with the Higgs boson.
Implications for phenomenology are briefly reviewed.

%[This is contribution to the Snowmass 2001 proceedings based
%on a talk given by the author.]

\end{abstract}
% \pacs{}

%\maketitle must follow title, authors, abstract and \pacs
\maketitle

%\label{}
%\subsection{}
%\subsubsection{}
\section{Introduction}

Randall and Sundrum (RS) proposed a fascinating solution to the hierarchy 
problem \cite{RS1}.  The setup involves two 4D surfaces (``branes'') 
bounding a slice of 5D compact AdS space taken to be on an $S^1/Z_2$ 
orbifold.  Gravity is effectively localized one
brane, while the Standard Model (SM) fields are 
assumed to be localized on the other.
The wavefunction overlap of the graviton with the SM brane is
exponentially suppressed, causing the masses of all fields localized
on the SM brane to be exponentially rescaled.  The hierarchy problem 
can be solved by assuming all fields initially have masses near the
4D Planck scale, and arranging that the exponential suppression rescales
the Planck mass to a TeV on the SM brane.  This requires 
stabilizing the size of the extra dimension to be about thirty-five
times larger than the AdS radius.  Goldberger and Wise \cite{GW}
proposed adding a massive bulk scalar field with suitable brane 
potentials causing it to acquire a vev with a nontrivial 
$x_5$-dependent profile.  The desired exponential suppression could be 
obtained without any large fine-tuning of parameters.
Fluctuations about the stabilized RS model include both tensor 
and scalar modes.  In this brief review, the fluctuations of the size 
of the extra dimension, characterized by the scalar ($g_{55}$) component 
of the metric otherwise known as the radion, are discussed.

\section{Radion mass}

Given a stabilizing mechanism, the radion 
mass can in principle be calculated.  In the Randall-Sundrum model with
a bulk scalar field, this was estimated in Refs.~\cite{GW,CGRT,GWpheno}.
A self-consistent calculation requires an ansatz of the scalar 
fluctuations about the RS background that solve the (linearized) Einstein 
equations \cite{CGR}, and incorporating
the backreaction of the bulk scalar vev into the metric \cite{DFGK}.
Treating the backreaction as a perturbation about the RS solution, 
the mass was found to be \cite{CGK} (see also \cite{TM})
\begin{equation}
m_r \sim (\mbox{backreaction}) \frac{\Lambda}{35}
\end{equation}
in terms of the ``warped'' Planck scale 
$\Lambda = e^{-k L} M_{\rm Pl} \sim {\cal O}({\rm TeV})$.
Note that in the limit the backreaction goes to zero, the radion mass
vanishes.  Thus, the expectation is that the radion is the lightest state 
in the RS model beyond the SM fields (as compared with the Kaluza-Klein 
gravitons that have masses of order $\Lambda$).  The precise mass of 
the radion is, however, dependent 
on the details of the stabilization mechanism.  For phenomenological 
purposes, a sensible range for the radion mass is probably 
${\cal O}(10 \; {\rm GeV}) \lesssim m_r \lesssim \Lambda$, 
where the lower limit arises from radiative corrections from SM fields, 
and the upper limit is the cutoff of the 4D effective theory.

\section{Radion couplings}

The linear coupling of the canonically normalized radion 
is \cite{CGRT,GWpheno}
\begin{equation}
- \frac{r(x)}{\sqrt{6} \Lambda} T^{\rm SM \, \mu}_\mu 
\label{P3-17-Kribs-couplings-eq}
\end{equation}
obtained by varying the action with respect to the $g_{55}$ component of
the metric.  Here $T^{\rm SM}_{\mu\nu}$ is the energy-momentum
tensor of the SM\@.  The key observation is that the coupling is
order $1/\Lambda$ (and not $1/M_{\rm Pl}$) as a result of the
exponential rescaling.  The radion can therefore be produced and decay 
on collider time scales.
The radion couples to all terms that violate scale invariance, which
can be easily read off from Eq.~(\ref{P3-17-Kribs-couplings-eq}).
At leading order this includes mass terms in the SM 
and kinetic terms for the Higgs boson (and Goldstone bosons).
This leads to 3-point couplings of the radion such as \\
\begin{picture}(400,80)
\qbezier(100,70)(104,74)(104,67.5)
\qbezier(104,67.5)(104,61)(108,65)
\qbezier(108,65)(112,69)(112,62.5)
\qbezier(112,62.5)(112,56)(116,60)
\qbezier(116,60)(120,64)(120,57.5)
\qbezier(120,57.5)(120,51)(124,55)
\qbezier(124,55)(128,59)(128,52.5)
\qbezier(128,52.5)(128,46)(132,50)
\qbezier(132,50)(136,54)(136,48.5)
\qbezier(136,48.5)(136,41)(140,45)
\qbezier(140,45)(144,49)(144,42.5)
\qbezier(144,42.5)(144,36)(148,40)
\qbezier(100,10)(104,6)(104,12.5)
\qbezier(104,12.5)(104,19)(108,15)
\qbezier(108,15)(112,11)(112,17.5)
\qbezier(112,17.5)(112,24)(116,20)
\qbezier(116,20)(120,16)(120,22.5)
\qbezier(120,22.5)(120,29)(124,25)
\qbezier(124,25)(128,21)(128,27.5)
\qbezier(128,27.5)(128,34)(132,30)
\qbezier(132,30)(136,26)(136,32.5)
\qbezier(136,32.5)(136,39)(140,35)
\qbezier(140,35)(144,31)(144,37.5)
\qbezier(144,37.5)(144,44)(148,40)
\put(148,40){\line(1,0){4}}
\put(156,40){\line(1,0){4}}
\put(164,40){\line(1,0){4}}
\put(172,40){\line(1,0){4}}
\put(180,40){\line(1,0){4}}
\put(188,40){\line(1,0){4}}
\put(196,40){\line(1,0){4}}
%
%\Photon(100,70)(150,40){4}{6}
%\Photon(100,10)(150,40){4}{6}
%\DashLine(150,40)(200,40){4}
\put(80,70){$W^+_\mu$}
\put(80,8){$W^-_\nu$}
\put(205,38){$r$}
\put(310,40){$-i \eta_{\mu\nu} \gamma 2 M_W^2/v$}
%\Photon(100,70)(150,40){4}{6}
%\Photon(100,10)(150,40){4}{6}
%\DashLine(150,40)(200,40){4}
%\Text(95,70)[r]{$W^+_\mu$}
%\Text(95,10)[r]{$W^-_\nu$}
%\Text(205,40)[l]{$r$}
%\Text(300,40)[l]{$-i \eta_{\mu\nu} \gamma 2 M_W^2/v$}
\end{picture} \\
where I have defined the coupling $\gamma \equiv v/(\sqrt{6} \Lambda)$. 
Since matter fermion and gauge boson masses arise through 
the Higgs vev, not surprisingly the radion couplings are 
rather similar to the Higgs boson.  Indeed tree-level couplings
of the radion can be obtained by replacing $h(x) \rightarrow -\gamma r(x)$
everywhere in the SM\@.
One important difference is that, due to the trace anomaly, 
there are direct couplings of the radion to massless gauge bosons 
at one-loop \cite{GRW,MD}.  For example, the radion coupling to 
two gluons through the trace anomaly is \\
\begin{picture}(400,80)
\qbezier(100,70)(104,74)(104,67.5)
\qbezier(104,67.5)(104,61)(108,65)
\qbezier(108,65)(112,69)(112,62.5)
\qbezier(112,62.5)(112,56)(116,60)
\qbezier(116,60)(120,64)(120,57.5)
\qbezier(120,57.5)(120,51)(124,55)
\qbezier(124,55)(128,59)(128,52.5)
\qbezier(128,52.5)(128,46)(132,50)
\qbezier(132,50)(136,54)(136,48.5)
\qbezier(136,48.5)(136,41)(140,45)
\qbezier(140,45)(144,49)(144,42.5)
\qbezier(144,42.5)(144,36)(148,40)
\qbezier(100,10)(104,6)(104,12.5)
\qbezier(104,12.5)(104,19)(108,15)
\qbezier(108,15)(112,11)(112,17.5)
\qbezier(112,17.5)(112,24)(116,20)
\qbezier(116,20)(120,16)(120,22.5)
\qbezier(120,22.5)(120,29)(124,25)
\qbezier(124,25)(128,21)(128,27.5)
\qbezier(128,27.5)(128,34)(132,30)
\qbezier(132,30)(136,26)(136,32.5)
\qbezier(136,32.5)(136,39)(140,35)
\qbezier(140,35)(144,31)(144,37.5)
\qbezier(144,37.5)(144,44)(148,40)
\put(148,40){\line(1,0){4}}
\put(156,40){\line(1,0){4}}
\put(164,40){\line(1,0){4}}
\put(172,40){\line(1,0){4}}
\put(180,40){\line(1,0){4}}
\put(188,40){\line(1,0){4}}
\put(196,40){\line(1,0){4}}
%
%\Gluon(100,70)(150,40){4}{6}
%\Gluon(100,10)(150,40){4}{6}
%\DashLine(150,40)(200,40){4}
\put(80,70){$g_\mu$}
\put(80,8){$g_\nu$}
\put(205,38){$r$}
\put(310,40){$-i [2 \, p \cdot q \, \eta_{\mu\nu} - p_\mu q_\nu - q_\mu p_\nu] 
              \gamma b_3 \frac{g_3^2}{32 \pi^2 v}$}
%\Text(95,70)[r]{$g_\mu$}
%\Text(95,10)[r]{$g_\nu$}
%\Text(205,40)[l]{$r$}
%\Text(300,40)[l]{$-i [2 p \cdot q \eta_{\mu\nu} - p_\mu q_\nu - q_\mu p_\nu] \gamma b_3 \frac{g_3^2}{32 \pi^2 v}$}
\end{picture} \\
where $b_3$ is the one-loop QCD $\beta$-function coefficient.

\section{Radion production}

The loop-suppressed but $\beta$-function enhanced coupling of the 
radion to gluons provides the dominant production process 
at hadron colliders (see e.g. \cite{GRW,MD,Cheung}).  At $e^+e^-$ 
machines, the dominant production processes are analogous to those 
of Higgs production, such as $e^+e^- \rightarrow Z r$.  In fact, the radion 
production cross section can be functionally related to the Higgs
production cross section
\begin{equation}
\sigma(e^+e^- \rightarrow Z r) = 
\gamma^2 \sigma(e^+e^- \rightarrow Z h; m_h \rightarrow m_r) \; .
\end{equation}
This leads to what are likely the best bounds on the radion mass
as function of $\gamma$ (or, equivalently, the cutoff scale).
A theoretical estimate of the bound was given in \cite{CGK}
(see also \cite{BKLL}), suggesting the radion mass is
unconstrained for $\gamma \lesssim 0.1$ ($\Lambda = 1$ TeV),
but must be $\gtrsim 100$ GeV for $\gamma \sim 1$.
In fact, an experimental analysis in the context of a two-Higgs 
doublet model, replacing $\sin^2(\beta-\alpha)$ with $\gamma^2$,
suggested this theoretical estimate was reasonable \cite{Krawczyk}.

\section{Radion decay}

The decay of the radion has been well-studied by e.g.\ \cite{GRW,Cheung}.
Due to the trace anomaly coupling to gluons, the dominant decay of 
the radion for masses between about $12 m_b \lesssim m_r \lesssim 2 M_W$
is $r \rightarrow gg$.  For somewhat lower masses, 
$r \rightarrow b\overline{b}$ is important, while for higher masses 
the decays $r \rightarrow W^+W^-, ZZ, hh$
have the largest branching ratios.  This means the search strategy 
for the radion could be quite different than that for the SM Higgs,
particularly for small to intermediate masses.    

\section{Radion-Higgs mixing}

Up to now I have assumed the sole interactions of the radion
are given by Eq.~(\ref{P3-17-Kribs-couplings-eq}).  
There is, however, one important
term in the low energy 4D effective theory allowed by all symmetries
that mixes the radion with the Higgs \cite{GRW}
\begin{equation}
S_{mixing} = \xi \int d^4 x \sqrt{g_{\rm ind}} R^{(4)}(g_{\rm ind}) H^\dag H 
\; ,
\end{equation}
where $R^{(4)}$ is the 4D Ricci scalar constructed out of the
induced metric on the SM brane, and $\xi$ is a dimensionless coupling.  
This leads to kinetic mixing 
between the radion and Higgs fields.  The physical mass eigenstates
are thus mixtures of the radion and Higgs; details can be found
in \cite{GRW,CGK}.

This can wreak havoc on your intuition concerning 
Higgs and radion fields.  Let me provide two examples. 
Precision electroweak observables receive contributions 
from both the radion and the Higgs.  In the absence of 
radion-Higgs mixing, the contribution of the radion is 
always small since it is functionally analogous to the Higgs, 
but with couplings suppressed by $\gamma$.  With significant radion-Higgs
mixing, the summed contribution of the two physical scalars (formerly
the radion and Higgs) can be quite different.  In particular,
it is possible to satisfy current constraints on the oblique
parameters S and T \cite{PT}
while pushing both the radion and Higgs masses to the several
hundred GeV range, provided $\xi$ is moderate and negative \cite{CGK}.
The second example is the partial-wave amplitude for electroweak
gauge boson scattering.  Again, in the absence of radion-Higgs mixing,
the radion contribution is always perturbative up to the 4D cutoff
scale $\Lambda$.  For large radion-Higgs mixing, however, the 
gauge boson scattering can become strong prior to reaching 
$\Lambda$ \cite{HKM}.  Other examples involving mixed scalar 
production and decay can be found in e.g.\ \cite{GRW,CDHY}.

\section{Summary}

I have sketched the mass, couplings, production, decay, 
and mixing of the radion in the Randall-Sundrum scenario.
The radion is expected to be the lightest state
beyond the SM with properties that are similar (but with
important differences) to the SM Higgs boson.  Furthermore, the 
radion and the Higgs can mix through an additional allowed
operator in the 4D effective theory.  If this mixing is significant, 
the direct and indirect effects of what were the physical radion and 
Higgs states can be significantly altered.

\begin{acknowledgments}

This summary is based on work done with Csaba Cs\'aki, 
Michael Graesser, Tao Han, and Bob McElrath, whom I thank 
for enjoyable collaborations and many discussions.  
At Snowmass I was supported in part by a DPF Snowmass Fellowship and 
by the U.S. Department of Energy under contract DE-FG02-95-ER40896.

\end{acknowledgments}

\bibliography{your bib file}

\end{document}